\documentclass[proceedings]{JHEP37}
\usepackage{amssymb}
\usepackage{amsfonts}
\usepackage{amsmath}
\usepackage{epsfig}

\newbox\mybox

\newcommand\fverb{\setbox\mybox=\hbox\bgroup\verb}
\newcommand\fverbdo{\egroup\medskip\noindent\fbox{\unhbox\mybox}\ }
\newcommand\fverbit{\egroup\item[\fbox{\unhbox\mybox}]}
\conference{E2-QES for non-Hermitian models}
\abstract{We propose the notion of $E_{2}$-quasi-exact solvability and apply this idea
to find explicit solutions to the eigenvalue problem for a non-Hermitian Hamiltonian system
depending on two parameters. The model considered reduces to the complex
Mathieu Hamiltonian in a double scaling limit, which enables us to compute
the exceptional points in the energy spectrum of the latter as a limiting process of the zeros for
some algebraic equations. The coefficient functions in the quasi-exact eigenfunctions
are univariate polynomials in the energy obeying a three-term recurrence relation. The latter
property guarantees the existence of a linear functional such that the
polynomials become orthogonal. The polynomials are shown to factorize for all levels above 
the quantization condition leading to vanishing norms rendering them to be weakly orthogonal.
In two concrete examples we compute the explicit expressions for the Stieltjes measure.}

\title{E2-quasi-exact solvability for non-Hermitian models}
\author{Andreas Fring \\
Department of Mathematics, City University London,\\
Northampton Square, London EC1V 0HB, UK\\
E-mail: a.fring@city.ac.uk}

\input{tcilatex}
\begin{document}

\section{Introduction}

In almost any study of quantum mechanical systems the determination of
explicit and ideally exact solutions to the associated eigenvalue problem is
either part of the key aims or at least essential ingredient. Integrability
is a property that largely facilitates to achieve this goal. Prime examples
for models that are integrable and can be solved exactly, classically as
well as quantum mechanically, are Calogero \cite{Cal2,Fring:2004zw} and
Sutherland models \cite{Suth3}. A systematic study of such type of models
can be established by relating them to Lie algebraic structures \cite%
{OP2,OP4} and exploiting the fact that the eigenfunctions of some
Hamiltonian systems form a flag which coincides with the finite dimensional
representation space of the associated Lie algebras. Models of this type for
which the representation space is finite are usually referred to as \emph{%
quasi-exactly solvable models} \cite{Turbiner00,Tur0} of Lie algebraic type.
Most of these models can be related to $sl_{2}(\mathbb{C})$ with their
compact and non-compact real forms $su(2)$ and $su(1,1)$, respectively \cite%
{Hum}.

However, in the context of extending \cite{PEGAAF2,Paulos} the study of such
models to non-Hermitian models of quasi-Hermitian/$\mathcal{PT}$-symmetric
type \cite{Urubu,Bender:1998ke,Benderrev,Alirev} it was recently found \cite%
{DFM} that a more natural setting for some systems is to cast them in terms
of Euclidean Lie algebras. Support for using the latter algebras in some
cases also comes from the actual solutions of the related eigenvalue
problem, i.e. the fact that the eigenfunctions of $sl_{2}(\mathbb{C})$
related systems are usually hypergeometric functions or their reduced
versions whereas the solutions of systems related Euclidean algebras are
usually Mathieu functions. It is well known that due to their different
singularity structure the two can not be related in the sense that one may
express Mathieu functions in terms of hypergeometric functions or vice
versa. Thus, systems of this type call for a different setting. Besides
these more mathematical aspects, a special interest in such type of models
results also from the fact that they are related to optical systems when
reducing the Schr\"{o}dinger equation to the Helmholtz equation. Concrete
versions of complex potentials leading to real Mathieu potentials have
recently been studied from a theoretical as well as experimental point of
view in \cite{Muss,MatMakris,Guo,OPMidya,MatHugh,MatHughEva,MatLongo}.

In the above spirit we suggest the general notion of $E_{2}$-quasi exact
solvability in a quite analogue fashion when compared to $sl_{2}(\mathbb{C})$%
-solvability. The key difference is that we cast the Hamiltonian in terms of
the generators of the Euclidean Lie algebra and also adapt the vector space
on which they act accordingly. We apply the general idea to solve a model,
which in a mildly modified version was first noted to be expressible in
terms of $sl_{2}(\mathbb{C})$-generators by Khare and Mandal \cite{KhareM}.
Making use of this quasi-exactly solvable structure the model was analyzed
further by Bagchi, Quesne Mallik and Roychoudhury in \cite{BijanMQR,BijanQR}
for the lowest levels. In particular the authors exploited the double
scaling limit towards the Mathieu equation and identified one exceptional
point. By reformulating the problem in terms of an $E_{2}$-setting we are
able to push this analysis further and provide a complete set of solutions
which in principle allows to identify all exceptional points from the double
scaling limit. We carry out the limit in two alternative ways, either on the
basis of the level-by-level solutions or directly for the recursive equation
itself and interpret the resulting equations as an eigenvalue problem for an
infinite matrix. The latter approach leads to much faster convergence.

We demonstrate that also at the heart of this type of $E_{2}$-quasi exact
solvability lies the fact the coefficient function polynomials in our
eigenfunctions possess the same essential features as the Bender-Dunne
polynomials \cite{Bender:1995rh}, i.e. they obey a three-term recurrence
relation which is reset at a certain level to a two-term relation with the
consequence that they factorize at the levels above. When going
\textquotedblleft on-shell\textquotedblright , i.e. taking the energy to be
quantized, one of the two factors vanishes. By Favard's theorem \cite{Favard}
the three-term recurrence relation already guarantees the existence of a
linear functional, such that the polynomials form an orthogonal set.
However, the fact that some of the polynomials vanish \textquotedblleft
on-shell\textquotedblright\ makes them weakly orthogonal polynomials as
defined in \cite{Finkel}. We determine the explicit form of the functional
by computing the expressions of the Stieltjes measure. For many
computations, such as the evaluation of the norm or the momentum functional,
the explicit knowledge of the measure is not needed when one assumes the
functional to exist and exploits the three-term recurrence relation.
However, this possibility is restricted to specific computations. Whenever
possible we use the two alternative ways to compute the same quantity to
demonstrate self-consistency.

\newpage

Our manuscript is organized as follows: Based on the Euclidean Lie algebra $%
E_{2}$ we propose in section 2 the notion of $E_{2}$-quasi-exact
solvability. In section 3 we apply this idea to solve the eigenvalue problem
for a non-Hermitian model that reduces to the complex Mathieu Hamiltonian in
the double scaling limit. We determine the exceptional points of the latter
in two alternative ways and provide a detailed analysis of the polynomial
coefficient functions with regard to their weakly orthogonal structure. Our
conclusions and a further outlook into open problems are stated in section
4.  

\section{$E_{2}$-quasi-exact solvability}

We introduce $E_{2}$-quasi-exactly solvable models in complete analogy to
the notion of $sl_{2}(\mathbb{C})$-quasi-exactly solvability originally
proposed by Turbiner \cite{Turbiner00,Tur0}. For this one considers
Hamiltonian operators $\mathcal{H}$ expressible in terms of $sl_{2}(\mathbb{C%
})$-generators acting on spaces $V_{n}$ of polynomials of order $n$ as $%
\mathcal{H}:V_{n}\mapsto V_{n}$ preserving the flag structure $V_{0}\subset
V_{1}\subset V_{2}\subset \ldots \subset V_{n}\subset \ldots $Whenever the
first eigenvalues and eigenfunctions of this sequence can be found the
models are referred to as \emph{quasi-exactly solvable}, whereas when the
entire infinite flag is preserved they are called \emph{exactly solvable}.
Integrability is conceptually different, but often related \cite{OP2,OP4}.
Here we express the Hamiltonians in terms of $E_{2}$-Lie algebraic
generators and proceed in a similar fashion.

The Euclidean algebra $E_{2}$ is the Lie algebra associated to the group $%
E(2)$ describing Euclidean transformations in the plane. We recall the
commutation relations obeyed by its three generators $u$, $v$ and $J$ 
\begin{equation}
\left[ u,J\right] =iv,\qquad \left[ v,J\right] =-iu,\qquad \text{and\qquad }%
\left[ u,v\right] =0.  \label{E2}
\end{equation}%
There are various useful representation for this algebra. We denote by $\Pi
^{(1)}$ the one acting on square integrable wavefunctions $L^{2}(\mathbf{S}%
^{1},d\theta )$ 
\begin{equation}
\Pi ^{(1)}:\quad J:=-i\partial _{\theta },\quad u:=\sin \theta ,\quad
v:=\cos \theta ,  \label{Rep1}
\end{equation}%
employed for instance in the context of quantizing strings on tori \cite%
{Isham}. The Casimir operator $C=u^{2}+v^{2}$ is here set to $1$. In
Cartesian coordinates the corresponding representations are for instance 
\begin{eqnarray}
\Pi ^{(2)} &:&\quad J:=yp_{x}-xp_{y},\quad u:=x,\quad ~v:=y,  \label{Rep2} \\
\Pi ^{(3)} &:&\quad J:=xp_{y}-p_{x}y,\quad u:=p_{y},\quad v:=p_{x},
\label{Rep3}
\end{eqnarray}%
where $x,y,p_{x},p_{y}$ are the Heisenberg canonical variables $%
x,y,p_{x},p_{y}$ with non-vanishing commutators $\left[ x,p_{x}\right] =%
\left[ y,p_{y}\right] =i$ in the convention $\hbar =1$. The constraint for (%
\ref{Rep2}) and (\ref{Rep3}) on the Casimir operator is viewed in coordinate
space as $x^{2}+y^{2}=1$ or momentum space as $p_{x}^{2}+p_{y}^{2}=1$,
respectively.

In \cite{DFM} five different types of anti-linear (or $\mathcal{PT}$)
symmetries for the $E_{2}$-algebra were identified. Adopting the notation
from there we are here especially interested in the symmetry $\mathcal{PT}%
_{3}:J\rightarrow J$, $u\rightarrow v$, $v\rightarrow u$, $i\rightarrow -i$,
as this one is also respected by the model considered below. For our
concrete representations this translates into $\mathcal{PT}_{3}^{\Pi
^{(1)}}:\theta \rightarrow \pi /2-\theta $, $i\rightarrow -i$, $\mathcal{PT}%
_{3}^{\Pi ^{(2)}}:x\rightarrow y$, $y\rightarrow x$, $p_{x}\rightarrow -p_{y}
$, $p_{y}\rightarrow -p_{x}$, $i\rightarrow -i$, and $\mathcal{PT}_{3}^{\Pi
^{(3)}}:x\rightarrow -y$, $y\rightarrow -x$, $p_{x}\rightarrow p_{y}$, $%
p_{y}\rightarrow p_{x}$, $i\rightarrow -i$. For the representation $\Pi
^{(1)}$ we now define two $\mathcal{PT}_{3}^{\Pi ^{(1)}}$-invariant vector
spaces over $\mathbb{R}$ as follows%
\begin{eqnarray}
V_{n}^{s}(\phi _{0}) &:&=\limfunc{span}\left\{ \left. \phi _{0}\left[ \sin
(2\theta ),i\sin (4\theta ),\ldots ,i^{n+1}\sin (2n\theta )\right]
\right\vert \theta \in \mathbb{R},\mathcal{PT}_{3}(\phi _{0})=\phi _{0}\in
L\right\} ,~~~~~~  \label{V1} \\
V_{n}^{c}(\phi _{0}) &:&=\limfunc{span}\left\{ \left. \phi _{0}\left[
1,i\cos (2\theta ),\ldots ,i^{n}\cos (2n\theta )\right] \right\vert \theta
\in \mathbb{R},\mathcal{PT}_{3}(\phi _{0})=\phi _{0}\in L\right\} .
\label{V2}
\end{eqnarray}%
Evidently we have $\mathcal{PT}_{3}^{\Pi ^{(1)}}:\psi \rightarrow \psi $ for
all $\psi \in V_{n}^{s,c}(\phi _{0})$. Next we compute the action of some $%
\mathcal{PT}_{3}$-invariant combinations of $E_{2}$-generators on these
spaces for concrete choices of $\phi _{0}$, which will turn out to be the
ground state wavefunction. Let us take $\phi _{0}^{c}=e^{i\kappa \cos
2\theta }$ and $\phi _{0}^{s}=e^{\kappa \sin 2\theta }$ with $\kappa \in 
\mathbb{R}$. Then we find%
\begin{eqnarray}
J &:&V_{n}^{s,c}\left( \phi _{0}^{c}\right) \mapsto V_{n+1}^{c,s}\left( \phi
_{0}^{c}\right) ,  \label{m1} \\
uv &:&V_{n}^{s,c}\left( \phi _{0}^{c}\right) \mapsto V_{n+1}^{c,s}\left(
\phi _{0}^{c}\right) ,  \label{m2} \\
i(u^{2}-v^{2}) &:&V_{n}^{s,c}\left( \phi _{0}^{c}\right) \mapsto
V_{n+1}^{s,c}\left( \phi _{0}^{c}\right) ,  \label{m3}
\end{eqnarray}%
and%
\begin{eqnarray}
J &:&V_{n}^{s,c}\left( \phi _{0}^{s}\right) \mapsto V_{n}^{c,s}\left( \phi
_{0}^{s}\right) \oplus V_{n+1}^{s,c}\left( \phi _{0}^{s}\right) ,  \label{b1}
\\
uv &:&V_{n}^{s,c}\left( \phi _{0}^{s}\right) \mapsto V_{n+1}^{c,s}\left(
\phi _{0}^{s}\right) ,  \label{b2} \\
i(u^{2}-v^{2}) &:&V_{n}^{s,c}\left( \phi _{0}^{s}\right) \mapsto
V_{n+1}^{s,c}\left( \phi _{0}^{s}\right) .  \label{b3}
\end{eqnarray}%
The actions of other invariant combinations, such as $i(v-u)$ and $v+u$, may
of course also be considered, but they do not map the spaces $%
V_{n}^{s,c}(\phi _{0}^{s,c})$ into each other. One may also include the
possibility to have odd integer coefficient for $\theta $ in the arguments
of the trigonometric functions in $V_{n}^{s,c}(\phi _{0})$. However, these
versions are more complicated and the two spaces start to mix when acted
upon with the $E_{2}$-generators.

Clearly this analysis should not be representation dependent and of course
one can transform $V_{n}^{s}(\phi _{0})$, $V_{n}^{c}(\phi _{0})$ and the
above arguments directly into the representations $\Pi ^{(2,3)}$. Proceeding
in a one-to-one fashion would involve a span over symmetric polynomials in $%
x,y$ or $p_{x},p_{y}$, which is not obvious to guess if one does not make a
reference to representation $\Pi ^{(1)}$. However, due to the constraint on
the Casimir operator one may trade all even powers in one variable for the
other and work with a simplified versions of polynomials in just one
variable. Thus for $\Pi ^{(2)}$ we define the vector spaces%
\begin{eqnarray}
\hat{V}_{n}^{s}(\phi _{0}) &:&=\limfunc{span}\left\{ \left. \phi _{0}\left(
xy,x^{3}y,\ldots ,x^{2n+1}y\right) \right\vert x\in \mathbb{R},\mathcal{PT}%
_{3}(\phi _{0})=\phi _{0}\in L\right\} ,~~~~~~ \\
\hat{V}_{n}^{c}(\phi _{0}) &:&=\limfunc{span}\left\{ \left. \phi _{0}\left(
1,x^{2},\ldots ,x^{2n}\right) \right\vert \theta \in \mathbb{R},\mathcal{PT}%
_{3}(\phi _{0})=\phi _{0}\in L\right\} .
\end{eqnarray}%
As a consequence of our replacements $y^{2n}\rightarrow (1-x^{2})^{n}$ we
have lost the explicit $\mathcal{PT}_{3}$-symmetry and must span the vector
space over $\mathbb{C}$ in this case. Taking now $\phi _{0}^{c}=e^{i\kappa
(y^{2}-x^{2})}$ and $\phi _{0}^{s}=e^{\kappa xy}$ with $\kappa \in \mathbb{R}
$ we find%
\begin{eqnarray}
J,uv &:&\hat{V}_{n}^{c}\left( \phi _{0}^{c}\right) \mapsto \hat{V}%
_{n}^{s}\left( \phi _{0}^{c}\right) , \\
J,uv &:&\hat{V}_{n}^{s}\left( \phi _{0}^{s}\right) \mapsto \hat{V}%
_{n+2}^{c}\left( \phi _{0}^{c}\right) , \\
i(u^{2}-v^{2}) &:&\hat{V}_{n}^{s,c}\left( \phi _{0}^{c}\right) \mapsto \hat{V%
}_{n+1}^{s,c}\left( \phi _{0}^{c}\right) ,
\end{eqnarray}%
and%
\begin{eqnarray}
J &:&\hat{V}_{n}^{s,c}\left( \phi _{0}^{s}\right) \mapsto \hat{V}%
_{n}^{s,c}\left( \phi _{0}^{s}\right) , \\
uv &:&\hat{V}_{n}^{c}\left( \phi _{0}^{s}\right) \mapsto \hat{V}%
_{n}^{s}\left( \phi _{0}^{s}\right) , \\
uv &:&\hat{V}_{n}^{s}\left( \phi _{0}^{s}\right) \mapsto \hat{V}%
_{n+2}^{c}\left( \phi _{0}^{s}\right) , \\
i(u^{2}-v^{2}) &:&\hat{V}_{n}^{s,c}\left( \phi _{0}^{s}\right) \mapsto \hat{V%
}_{n+1}^{s,c}\left( \phi _{0}^{s}\right) .
\end{eqnarray}%
Analogues of representation $\Pi ^{(2)}$ will allow for easy extension of
these arguments to higher rank algebras. Similarly one might set up a
vectorspace for the representation $\Pi ^{(3)}$.

\section{Complex Mathieu equation from a large N-limit}

Let us now consider the $\mathcal{PT}_{3}$-symmetric non-Hermitian
Hamiltonian%
\begin{equation}
\mathcal{H}_{\text{Mat}}=J^{2}+2ig(u^{2}-v^{2}).
\end{equation}%
For the representation $\Pi ^{(1)}$ the Hamiltonian $\mathcal{H}_{\text{Mat}}
$ evidently becomes \cite{DFM}%
\begin{equation}
\mathcal{H}_{\text{Mat}}^{\Pi ^{(1)}}=-\frac{d^{2}}{d\theta ^{2}}+2ig\cos
(2\theta ),  \label{HMat}
\end{equation}%
whose corresponding time-independent Schr\"{o}dinger equation $\mathcal{H}_{%
\text{Mat}}^{\Pi ^{(1)}}\psi =E\psi $ is the complex Mathieu equation solved
by%
\begin{equation}
\psi (\theta )=c_{1}C\left( E,ig,\theta \right) +c_{2}S\left( E,ig,\theta
\right) ,
\end{equation}%
with $C$ and $S$ denoting the even and odd Mathieu function, respectively.
From (\ref{m1}) and (\ref{m3}) we observe that $\mathcal{H}_{\text{Mat}%
}^{\Pi ^{(1)}}:V_{n}^{s,c}\left( \phi _{0}^{c}\right) \mapsto
V_{n+2}^{s,c}\left( \phi _{0}^{c}\right) \oplus V_{n+1}^{s,c}\left( \phi
_{0}^{c}\right) $. Thus in order to solve the eigenvalue problem even at the
lowest level we require two constraints to reduce the dimension. It is clear
that this is not possible, since there is for instance only one term at
level $n+2$ such that it can not be compensated for by any counterterms.
Using $\phi _{0}^{s}$ or other $\mathcal{PT}_{3}$-invariant functions
instead of $\phi _{0}^{c}$ does not improve the scenario. Thus the
Hamiltonian $\mathcal{H}_{\text{Mat}}$ does not correspond to a
quasi-exactly solvable $E_{2}$-system in the sense defined above.

However, considering instead the $\mathcal{PT}_{3}$-symmetric non-Hermitian
Hamiltonian 
\begin{equation}
\mathcal{H}_{N}=J^{2}+\zeta ^{2}(u^{2}-v^{2})^{2}+2i\zeta N(u^{2}-v^{2}),
\label{HN}
\end{equation}%
we obtain $\mathcal{H}_{N}:V_{n}^{s,c}\left( \phi _{0}^{c}\right) \mapsto
V_{n+2}^{s,c}\left( \phi _{0}^{c}\right) \oplus \zeta
^{2}V_{n+2}^{s,c}\left( \phi _{0}^{c}\right) \oplus V_{n+1}^{s,c}\left( \phi
_{0}^{c}\right) $. Now quasi-exact solvability is achievable since we have
the possibility to reduce the dimension of $V_{n+2}^{s,c}\left( \phi
_{0}^{c}\right) \oplus \zeta ^{2}V_{n+2}^{s,c}\left( \phi _{0}^{c}\right) $
by one and impose a further constraint at the level $n+1$. Indeed taking $%
\kappa =\zeta /2$ in $\phi _{0}^{c}$ we find $\mathcal{H}%
_{N}:V_{(N-1)/2}^{s,c}\left( \phi _{0}^{c}\right) \mapsto
V_{(N-1)/2}^{s,c}\left( \phi _{0}^{c}\right) $.

We further note that in a double scaling limit the Hamiltonian $\mathcal{H}%
_{N}$ converges to the complex Mathieu Hamiltonian 
\begin{equation}
\lim_{N\rightarrow \infty ,\zeta \rightarrow 0}\mathcal{H}_{N}=\mathcal{H}_{%
\text{Mat}},\qquad \text{for }g:=N\zeta <\infty .
\end{equation}

Thus by studying properties of the quasi-solvable model $\mathcal{H}_{N}$ we
can obtain non-trivial information about the system related to $\mathcal{H}_{%
\text{Mat}}$ in the double scaling limit. This idea was first pursued in 
\cite{BijanMQR,BijanQR} for a slightly shifted system. One of the main
differences here is that $\mathcal{H}_{N}$ and $\mathcal{H}_{\text{Mat}}$
are now expressed in terms of $E_{2}$-Lie algebraic generators rather than $%
sl_{2}(\mathbb{C})$-generators, i.e. we exploit $E_{2}$-quasi-exact
solvability instead of the standard one. We will also extract more
information than in \cite{BijanMQR,BijanQR}.

\subsection{Exact eigenfunctions from three-term recurrence relations}

Let us now see in detail how to solve the Schr\"{o}dinger equation $\mathcal{%
H}_{N}\psi _{N}=E_{N}\psi _{N}$ level by level and in closed form. In
accordance with the above observation that $\mathcal{H}%
_{N}:V_{(N-1)/2}^{s,c}\left( \phi _{0}^{c}\right) \mapsto
V_{(N-1)/2}^{s,c}\left( \phi _{0}^{c}\right) $, we take the following Ansatz
for our eigenfunctions%
\begin{eqnarray}
\psi _{N}^{s}(\theta ) &=&ie^{\frac{i}{2}\zeta \cos (2\theta
)}\sum_{n=1}^{(N-1)/2}i^{n}\sigma _{n}\sin (2n\theta )\in
V_{(N-1)/2}^{s}(\phi _{0}^{c}),\quad \text{\quad }  \label{psis} \\
\psi _{N}^{c}(\theta ) &=&e^{\frac{i}{2}\zeta \cos (2\theta
)}\sum_{n=0}^{(N-1)/2}i^{n}c_{n}\cos (2n\theta )\in V_{(N-1)/2}^{c}(\phi
_{0}^{c}),  \label{psic}
\end{eqnarray}%
with $N$ being odd and the unknown functions $\sigma _{n}$ and $c_{n}$ to be
determined. One may of course also take $N$ to be even, but it turns out
that the related spectra are complex throughout the entire range of $\zeta $
and those solutions are therefore less interesting. The upper limit in the
sums may appear somewhat artificial at this point and in fact one could
consider the entire Fourier series. However, as we shall see in detail below
the series truncate automatically at the stated level when going
\textquotedblleft on-shell\textquotedblright\ due to the factorization of
the functions $\sigma _{n}$ and $c_{n}$. This is the same mechanism as
originally observed by Bender and Dunne in \cite{Bender:1995rh} for $sl_{2}(%
\mathbb{C})$ solvable models. Notice also that for $\sigma _{n},c_{n}\in 
\mathbb{R}$ the eigenfunctions are designed to be manifestly $\mathcal{PT}%
_{3}$-symmetric.

Upon substituting $\psi _{N}^{s}$ into the Schr\"{o}dinger equation we find
the three-term recurrence equations%
\begin{equation}
\zeta (2n-N-1)\sigma _{n-1}+\left( 4n^{2}+\zeta ^{2}-E\right) \sigma
_{n}+\zeta (2n+1+N)\sigma _{n+1}=0,  \label{s1}
\end{equation}%
for $\sigma _{n}$ with $n=1,\ldots ,(N-1)/2$ by reading off the coefficients
of $\sin (2n\theta )$. These equations may be solved successively by making
a suitable assumption about the initial condition. Taking therefore $\sigma
_{1}=1$ we easily solve (\ref{s1}) with $n=1$ for $\sigma _{2}$.
Subsequently we find $\sigma _{3}$ from (\ref{s1}) with $n=2$, $\sigma _{4}$
from (\ref{s1}) with $n=3$, etc. until we reach the equation (\ref{s1}) with 
$n=(N-1)/2$%
\begin{equation}
2\zeta \sigma _{\frac{N-3}{2}}+\left[ E-\zeta ^{2}-(N-1)^{2}\right] \sigma _{%
\frac{N-1}{2}}=0.  \label{QC}
\end{equation}%
Since there are no free coefficient functions $\sigma _{n}$ left, this
equation needs to be consistent in itself. This can indeed be achieved when
interpreting it as the quantization condition for the energies $E_{N}^{s}$
as discussed below. Proceeding in this manner we obtain the coefficient
functions 
\begin{eqnarray}
\sigma _{2} &=&\frac{E-\zeta ^{2}-4}{\zeta (N+3)}, \\
\sigma _{3} &=&\frac{E^{2}-2E\left( \zeta ^{2}+10\right) +\zeta ^{4}+\zeta
^{2}\left( N^{2}+11\right) +64}{\zeta ^{2}(N+3)(N+5)}, \\
\sigma _{4} &=&\frac{E^{3}-E^{2}\left( 3\zeta ^{2}+56\right) +E\left[ 3\zeta
^{4}+2\zeta ^{2}\left( N^{2}+39\right) +784\right] }{\zeta
^{3}(N+3)(N+5)(N+7)} \\
&&-\frac{\zeta ^{6}+2\zeta ^{4}\left( N^{2}+11\right) +40\zeta ^{2}\left(
N^{2}+9\right) +2304}{\zeta ^{3}(N+3)(N+5)(N+7)}.  \notag
\end{eqnarray}%
Expressions for $\sigma _{n}$ for larger values of $n$ can easily be
obtained, but evidently they\ are rather lengthy and therefore not reported
here in explicit form. Following \cite{Gonoskov} we may in fact solve the
three-term recurrence equations (\ref{s1}) in closed form. For this purpose
we first convert (\ref{s1}) into the canonical form for three-term
recurrence relations%
\begin{equation}
s_{n+1}=s_{n}+\gamma _{n}s_{n-1},  \label{canform}
\end{equation}%
by means of the relations%
\begin{equation}
\sigma _{n}=s_{n-1}\prod\limits_{k=1}^{n-1}\alpha _{k},\quad \alpha _{n}:=%
\frac{E-\zeta ^{2}-4n^{2}}{\zeta (2n+N+1)},\quad \beta _{n}:=\frac{N+1-2n}{%
N+1+2n},\quad \gamma _{n}:=\frac{\beta _{n+1}}{\alpha _{n+1}\alpha _{n}}.
\label{abc}
\end{equation}%
A generic solution for (\ref{canform}) in closed algebraic form was derived
by Gonoskov in \cite{Gonoskov} 
\begin{equation}
s_{n}=1+\sum\limits_{p=1}^{\left\lfloor n/2\right\rfloor }S(n-1,p),
\label{clo}
\end{equation}%
where $\left\lfloor x\right\rfloor :=\max (n\in \mathbb{Z}|n\leq x)$ and $%
S(n,p)$ is a $p$-fold sum%
\begin{equation}
S(n,p):=\sum\limits_{k_{1}=1}^{n+2(1-p)}\gamma
_{k_{1}}\sum\limits_{k_{2}=2+k_{1}}^{n+2(2-p)}\gamma _{k_{2}}\ldots
\sum\limits_{k_{p-1}=2+k_{p-2}}^{n-2}\gamma
_{k_{p-1}}\sum\limits_{k_{2}=2+k_{p-1}}^{n}\gamma _{k_{p}}.
\end{equation}%
Taking $s_{0}=1$, $s_{1}=1$ the expression (\ref{clo}) yields $%
s_{2}=1+\gamma _{1}$, $s_{3}=1+\gamma _{1}+\gamma _{2}$, $s_{4}=1+\gamma
_{1}+\gamma _{2}+\gamma _{3}+\gamma _{1}\gamma _{3}$, $s_{5}=1+\gamma
_{1}+\gamma _{2}+\gamma _{3}+\gamma _{4}+\gamma _{1}\gamma _{3}+\gamma
_{1}\gamma _{4}+\gamma _{2}\gamma _{4}$, etc. Clearly (\ref{clo}) is not of
a compact closed form one expects for instance from a three-term recurrence
relation with constant coefficient as it still involves a $p$-fold sum.
Nonetheless, these sums can be computed easily.

We observe here that (\ref{clo}) admits an interesting alternative
representation related to a generalized version of Pascal's triangle. To see
this we define first recursively the sets%
\begin{equation}
M_{ij}=M_{(i-1)j}\cup \gamma _{i+j-1}M_{(i-1)(j-1)}\quad \text{for }i,j\in 
\mathbb{N}\text{, }i\geq j\text{, }M_{00}=\{1\},
\end{equation}%
such that for instance $M_{10}=M_{00}=\{1\}$, $M_{11}=\gamma
_{1}M_{00}=\{\gamma _{1}\}$, $M_{20}=M_{10}=\{1\}$, $M_{21}=M_{11}\cup
\gamma _{2}M_{10}=\{\gamma _{1},\gamma _{2}\}$, $M_{22}=\gamma
_{3}M_{11}=\{\gamma _{1}\gamma _{3}\}$,$\ldots $ As depicted in figure \ref%
{Pas} Pascal's triangle is generated by producing the element in row $i$ and
column $j$ as the union of the two sets in the previous row $i-1$ where the
one to the left is multiplied by a value $\gamma _{k}$ as specified.

\begin{figure}[h]
\begin{center}
\includegraphics[width=12cm]{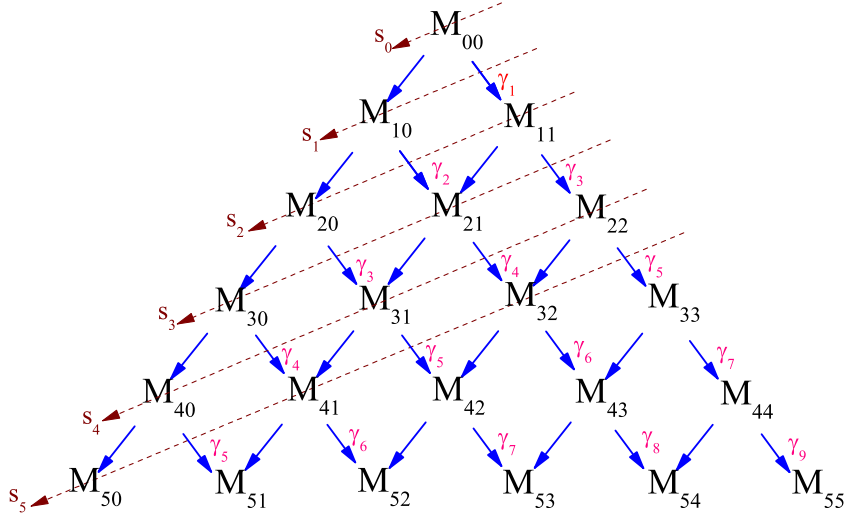}
\end{center}
\caption{Solutions to the canonical three-term recurrence relation (\protect
\ref{canform}) from shallow unions in a generalized Pascal's triangle.}
\label{Pas}
\end{figure}
\noindent From these sets we form the \textquotedblleft shallow
unions\textquotedblright\ 
\begin{equation}
\mathcal{M}_{n}=\bigcup\limits_{k=0}^{\left\lfloor (n+1)/2\right\rfloor
}M_{(n-k)k},
\end{equation}%
in analogy to the shallow sum in a standard Pascal's triangle. Then the
solution for the canonical three-term recurrence relation (\ref{canform})
can be expressed as%
\begin{equation}
s_{n}=\sum\limits_{k=1}^{F_{n+1}}x_{k}\qquad \text{for }x_{k}\in \mathcal{M}%
_{n}\text{,}  \label{sn}
\end{equation}%
where the number of elements in $\mathcal{M}_{n}$ equals the n-th Fibonacci
number $F_{n}$, i.e. $F_{1}=1$, $F_{2}=1$, $F_{3}=2$, $F_{4}=3$, $F_{5}=5$, $%
F_{8}=8$,$\ldots $ The occurrence of the Fibonacci numbers is to be expected
since for $\gamma _{n}=1$ the canonical relation (\ref{canform}) simply
reduces to the well-known Fibonacci~three-term recurrence relation and in
this case (\ref{sn}) is just counting the elements in $\mathcal{M}_{n}$.

For the second Ansatz $\psi _{N}^{c}$ in (\ref{psic}) we may proceed in a
similar fashion. Reading off the coefficients of $\cos (2n\theta )$ from the
result of the substitution into the Schr\"{o}dinger equation yields the same
recursive equations (\ref{s1}) with $\sigma _{n}$ replaced by $c_{n}$ 
\begin{equation}
\zeta (2n-N-1)c_{n-1}+\left( 4n^{2}+\zeta ^{2}-E\right) c_{n}+\zeta
(2n+1+N)c_{n+1}=0,  \label{c2}
\end{equation}%
albeit only valid for $n=0,2,\ldots ,(N-1)/2$. In addition we have the
relation 
\begin{equation}
\mathbf{2}\zeta (1-N)c_{0}+\left( 4+\zeta ^{2}-E\right) c_{1}+\zeta
(3+N)c_{2}=0,  \label{c3}
\end{equation}%
which is not part of the general family of equations (\ref{c2}) due to the
factor 2 in the first term. Taking $c_{0}=1$ we solve (\ref{c2}) with $n=0$
for $c_{1}$, (\ref{c3}) for $c_{2}$, (\ref{c2}) with $n=2$ for $c_{3}$,
until we reach the quantization condition at $n=(N-1)/2$. In this way we
find 
\begin{eqnarray}
c_{1} &=&\frac{E-\zeta ^{2}}{\zeta (N+1)}, \\
c_{2} &=&\frac{E^{2}-2E\left( \zeta ^{2}+2\right) }{\zeta ^{2}(N+1)(N+3)}+%
\frac{\zeta ^{2}+2N^{2}+2}{(N+1)(N+3)}, \\
c_{3} &=&\frac{E^{3}-E^{2}\left( 3\zeta ^{2}+20\right) +E\left[ 3\zeta
^{4}+\zeta ^{2}\left( 3N^{2}+29\right) +64\right] }{\zeta ^{3}(N+1)(N+3)(N+5)%
} \\
&&-\frac{\zeta ^{4}+9\zeta ^{2}+\left( 3\zeta ^{2}+32\right) N^{2}+32}{\zeta
(N+1)(N+3)(N+5)}.  \notag
\end{eqnarray}%
Again these recurrence relations may be solved in closed form. We convert (%
\ref{c2}) and (\ref{c3}) into the canonical three-term relation form%
\begin{equation}
q_{n+1}=q_{n}+\lambda _{n}q_{n-1},  \label{recq}
\end{equation}%
by means of the substitutions%
\begin{equation}
c_{n}=q_{n}\prod\limits_{k=0}^{n-1}\alpha _{k},\quad \lambda _{1}=\frac{%
2\beta _{1}}{\alpha _{1}\alpha _{0}},\quad \lambda _{\ell }=\frac{\beta
_{\ell }}{\alpha _{\ell }\alpha _{\ell -1}},\quad \text{for }\ell \in 
\mathbb{N}_{0}\backslash \{1\}.  \label{cl}
\end{equation}%
The functions $\alpha _{n}$ and $\beta _{n}$ are defined as in equation (\ref%
{abc}). Taking $q_{0}=1$, $q_{1}=1$, i.e. $c_{0}=q_{0}=1$, $%
c_{1}=q_{1}\alpha _{0}=\alpha _{0}$ the expression (\ref{clo}) yields a
closed solution by replacing $\sigma _{n}\rightarrow c_{n}$ and $\gamma
_{n}\rightarrow \lambda _{n}$, e.g. $c_{2}=q_{2}\alpha _{0}\alpha
_{1}=(1+\lambda _{1})\alpha _{0}\alpha _{1}=\alpha _{0}\alpha _{1}+2\beta
_{1}$, $c_{3}=q_{3}\alpha _{0}\alpha _{1}\alpha _{2}=(1+\lambda _{1}+\lambda
_{2})\alpha _{0}\alpha _{1}\alpha _{2}=\alpha _{0}\alpha _{1}\alpha
_{2}+2\beta _{1}\alpha _{2}+\beta _{2}\alpha _{0}$, etc.

\subsection{Energy quantization and the double scaling limit}

\begin{figure}[h]
\centering   \includegraphics[width=7.5cm,height=6.0cm]{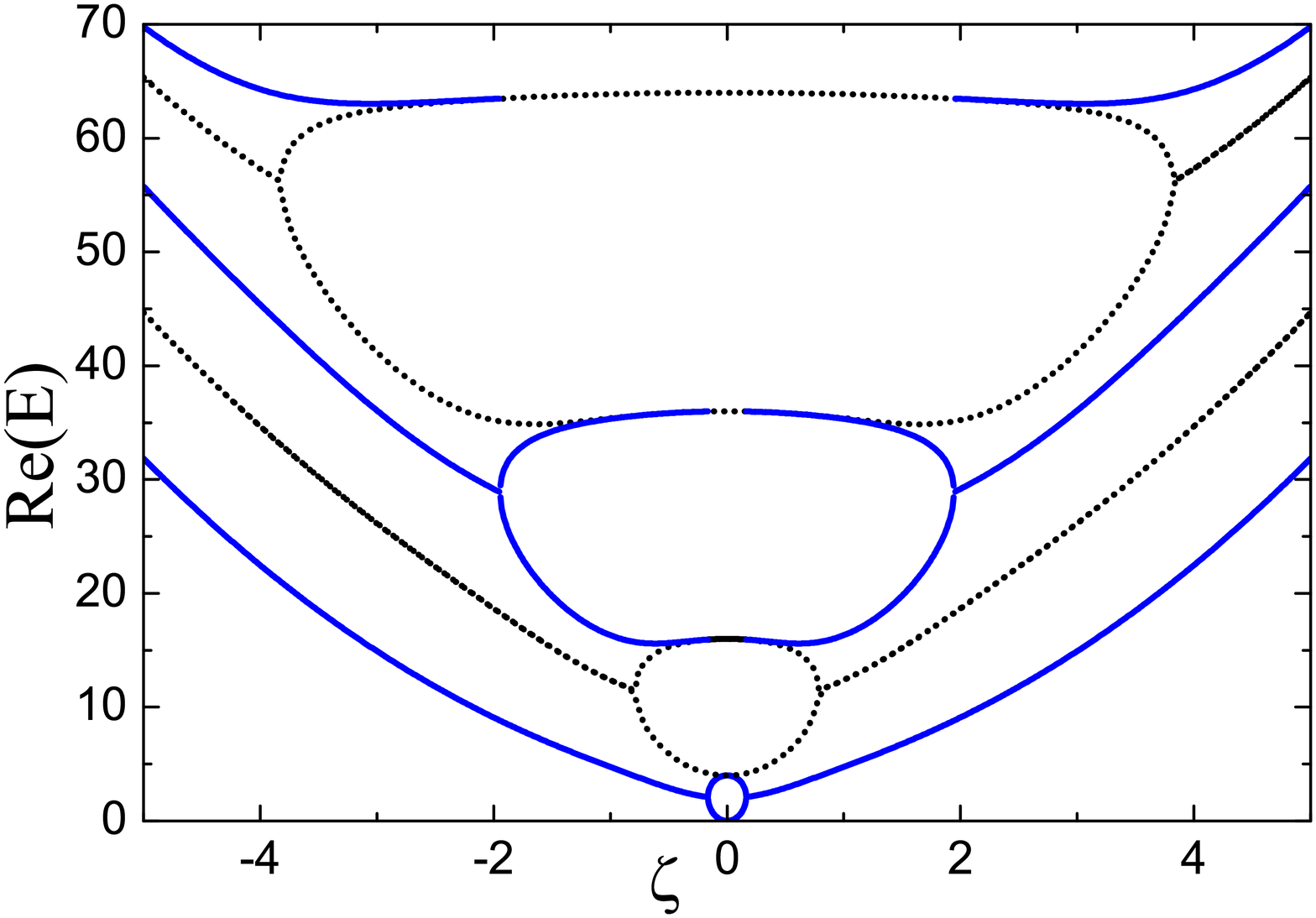} %
\includegraphics[width=7.5cm,height=6.0cm]{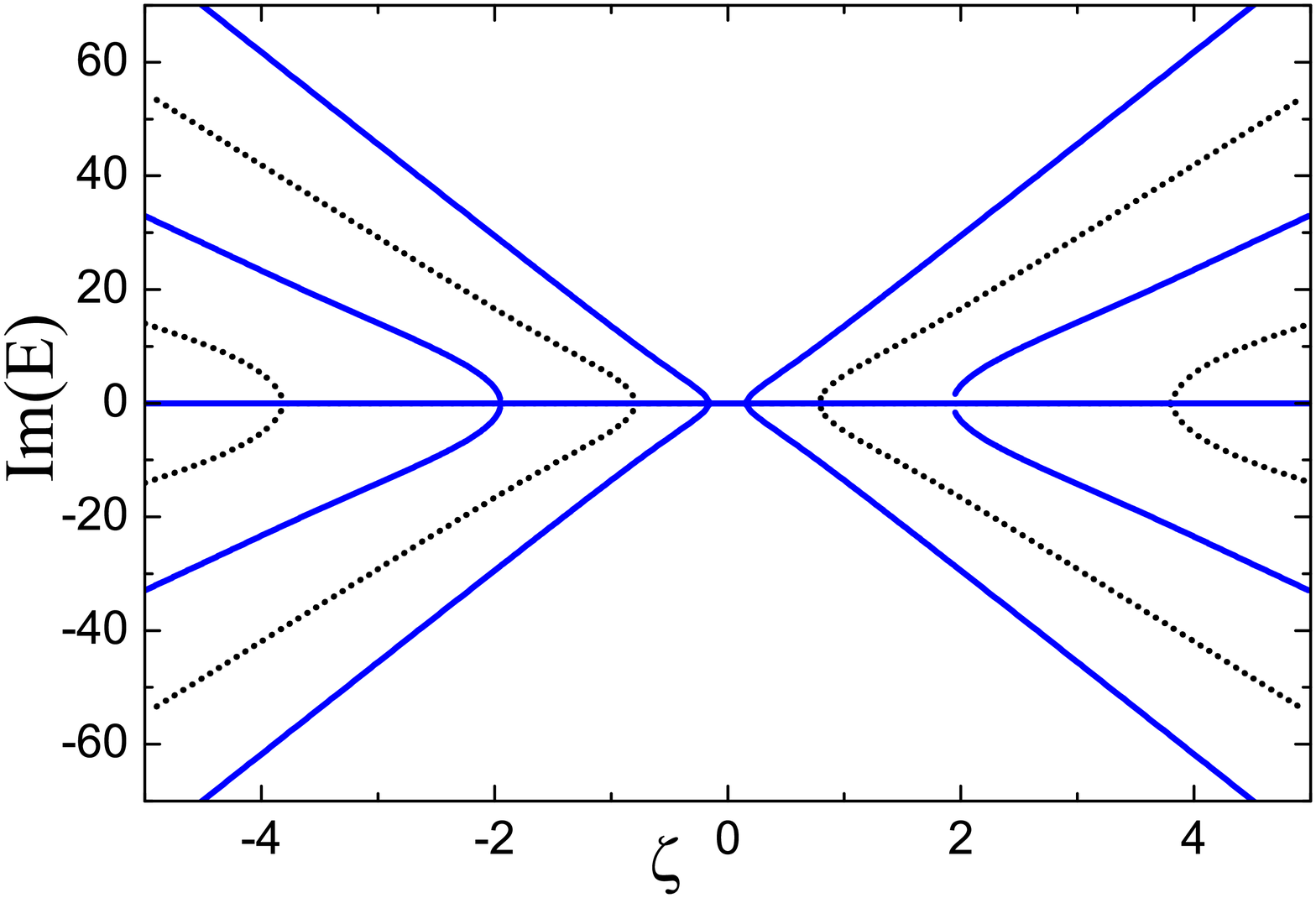} \centering   
\caption{Energy eigenvalues $E_{9}$ as a function of $\protect\zeta $
associated to odd (dotted, black) and even (solid, blue) eigenfunctions.}
\label{fig2}
\end{figure}

As mentioned, in the last step of the solution procedure for the recurrence
relations we have no free coefficient function left. We can, however, fix
the energy to a specific value, such that the last equation constitutes the
energy quantization condition. We simple have to solve the algebraic
expressions%
\begin{equation}
\sigma _{(N+1)/2}=s_{(N-1)/2}\prod\limits_{k=1}^{(N-1)/2}\alpha _{k}=0\qquad 
\text{and\qquad }c_{(N+1)/2}=q_{(N+1)/2}\prod\limits_{k=0}^{(N-1)/2}\alpha
_{k}=0,  \label{EQ}
\end{equation}%
for $E_{N}$ at specific values of $N$. We focus here on $N$ being odd as
these are the interesting cases producing spectra that, depending on the
values of $\zeta $, are real in part with fully unbroken $\mathcal{PT}$%
-symmetry and also possess spontaneously broken phases with energies
occurring in complex conjugate pairs. For even values of $N$ the $\mathcal{PT%
}$-symmetry is broken throughout the entire range of $\zeta $. From (\ref{EQ}%
) we observe that the degree of the polynomial in the energies $E_{N}$
equals the number of factors $\alpha _{k}$ in the product. Thus for the
eigenfunctions related to the Ansatz (\ref{psis}) we find the corresponding
energy eigenvalues $E_{N}^{s}$ from solving the first equation in (\ref{EQ}%
), that is a $(N-1)/2$-th order equations in $E$, to 
\begin{eqnarray}
E_{3}^{s} &=&4+\zeta ^{2}, \\
E_{5}^{s,\pm } &=&10+\zeta ^{2}\pm 2\sqrt{9-4\zeta ^{2}}, \\
E_{7}^{s,\ell } &=&\frac{56}{3}+\zeta ^{2}+\frac{4}{3}e^{\frac{i\pi \ell }{3}%
}\Omega +\frac{4}{3}e^{-\frac{i\pi \ell }{3}}\Omega ^{-1}\left( 49-12\zeta
^{2}\right) ,\qquad ~~  \label{E7}
\end{eqnarray}%
with $\ell =0,\pm 2$ and $\Omega :=\left[ 143+72\zeta ^{2}+6\sqrt{3}\sqrt{%
16\zeta ^{6}-148\zeta ^{4}+991\zeta ^{2}-900}\right] ^{1/3}$. For the
solutions related to (\ref{psic}) we compute the eigenenergies $E_{N}^{s}$
by solving the first equation in (\ref{EQ}), that is a $(N+1)/2$-th order
equation in $E$, to%
\begin{eqnarray}
E_{1}^{c} &=&\zeta ^{2}, \\
E_{3}^{c,\pm } &=&2+\zeta ^{2}\pm 2\sqrt{1-4\zeta ^{2}}, \\
E_{5}^{c,\ell } &=&\frac{20}{3}+\zeta ^{2}+\frac{4}{3}e^{\frac{i\pi \ell }{3}%
}\hat{\Omega}+\frac{4}{3}e^{-\frac{i\pi \ell }{3}}\hat{\Omega}^{-1}\left(
13-12\zeta ^{2}\right) ,\qquad ~~  \label{E5}
\end{eqnarray}%
with $\ell =0,\pm 2$ and $\hat{\Omega}:=\left[ 35+72\zeta ^{2}+6\sqrt{3}%
\sqrt{16\zeta ^{6}-4\zeta ^{4}+103\zeta ^{2}-9}\right] ^{1/3}$. At higher
levels the expressions become increasingly complex and are therefore not
reported here. However, using the generic solutions above they may be
obtained easily in a numerical form, being just limited by computer power.
As an example we depict the values for $N=9$ in figure \ref{fig2}.

We observe the typical coalescence of two levels at the transition from the $%
\mathcal{PT}$-symmetric to the spontaneously broken regime marked by the
so-called exceptional points \cite{Kato,HeissEx,IngridEx,IngridUwe}. The
precise values can be found by determining the real roots of the
discriminants of the polynomials (\ref{EQ}) in $E$ at the quantization
level. Introducing the coefficients $a_{i}$ via the expansion for the
polynomials by $P(E)=\sum\nolimits_{k=0}^{n}a_{k}E^{k}$, the discriminants
can be computed easily from the determinants of the Sylvester matrix $S$
formed from $P(E)$ and its derivative $dP(E)/dE$ with entries%
\begin{equation}
S_{ij}=\left\{ 
\begin{array}{ll}
a_{n+i-j}, & \text{for }1\leq i\leq n-1,1\leq j\leq 2n-1, \\ 
(1+i-j)a_{1+i-j},\quad  & \text{for }n\leq i\leq 2n-1,1\leq j\leq 2n-1,%
\end{array}%
\right.   \label{Sylvester}
\end{equation}%
see for instance \cite{Akritas}. The discriminants are then given by $\det
S=\kappa _{N}\Delta _{N}$. Dropping the overall factors $\kappa _{N}$ that
do not contribute to the zeros we compute in this way

\begin{eqnarray}
\Delta _{1}^{s} &=&\Delta _{1}^{c}=\Delta _{3}^{s}=1, \\
\Delta _{3}^{c} &=&2^{2}\zeta ^{2}-1,  \notag \\
\Delta _{5}^{s} &=&2^{2}\zeta ^{2}-9,  \notag \\
\Delta _{5}^{c} &=&2^{4}\zeta ^{6}-4\zeta ^{4}+103\zeta ^{2}-9,  \notag \\
\Delta _{7}^{s} &=&2^{4}\zeta ^{6}-148\zeta ^{4}+991\zeta ^{2}-900,  \notag
\\
\Delta _{7}^{c} &=&2^{8}\zeta ^{12}+128\zeta ^{10}+6160\zeta
^{8}-132480\zeta ^{6}+16128\zeta ^{4}-183069\zeta ^{2}+8100,  \notag \\
\Delta _{9}^{s} &=&2^{8}\zeta ^{12}-6016\zeta ^{10}+112656\zeta
^{8}-1526400\zeta ^{6}+6645600\zeta ^{4}-19579725\zeta ^{2}+9922500~~,~ 
\notag \\
\Delta _{9}^{c} &=&2^{14}\zeta ^{20}+40960\zeta ^{18}+805888\zeta
^{16}-50754048\zeta ^{14}+1392155136\zeta ^{12}+290722752\zeta ^{10}  \notag
\\
&&+11896163064\zeta ^{8}-113625663975\zeta ^{6}+8084618100\zeta
^{4}-53543300400\zeta ^{2}+1428840000,  \notag \\
\Delta _{11}^{s} &=&2^{14}\zeta ^{20}-778240\zeta ^{18}+29068288\zeta
^{16}-922894848\zeta ^{14}+21572729856\zeta ^{12}  \notag \\
&&-252561878208\zeta ^{10}+2351098873944\zeta ^{8}-15196772229975\zeta
^{6}+39898457932725\zeta ^{4}  \notag \\
&&-69457284086400\zeta ^{2}+22684263840000  \notag \\
\Delta _{11}^{c} &=&2^{22}\zeta ^{30}+26214400\zeta ^{28}+274464768\zeta
^{26}-43646517248\zeta ^{24}+2888883798016\zeta ^{22}\ldots  \notag
\end{eqnarray}%
The positive real roots $\zeta _{0}$ of these polynomials multiplied by $N$
are reported in table \ref{T1}.

\begin{table}[h]
\begin{center}
\begin{tabular}{c||c|c|c|c|c|}
$N$ & $\zeta _{0}N$ & $\zeta _{0}N$ & $\zeta _{0}N$ & $\zeta _{0}N$ & $\zeta
_{0}N$ \\ \hline\hline
$3$ & $1.50000$ &  &  &  &  \\ \hline
$5$ & $1.47963$ & $7.50000$ &  &  &  \\ \hline
$7$ & $1.47426$ & $7.19195$ & $18.4246$ &  &  \\ \hline
$9$ & $1.47208$ & $7.08219$ & $17.5098$ & $34.4001$ &  \\ \hline
$11$ & $1.47098$ & $7.02966$ & $17.1292$ & $32.5974$ & $55.4904$ \\ \hline
& $\vdots $ & $\vdots $ & $\vdots $ & $\vdots $ & $\vdots $ \\ \hline
$\infty $ & $1.46877$ & $6.92895$ & $16.4711$ & $30.0967$ & $47.806$%
\end{tabular}%
\end{center}
\caption{Values of $\protect\zeta _{0}N$ computed from the positive real
zeros $\protect\zeta _{0}$ of the discriminant polynomials $\Delta _{N}(%
\protect\zeta )$ converging to the critical values obtained from a Floquet
method.}
\label{T1}
\end{table}

We observe that the product $\zeta _{0}N$ appears to converges to the
critical values of the complex Mathieu equation energy eigenvalues related
to the non-Hermitian Hamiltonian $\mathcal{H}_{\text{Mat}}^{\Pi ^{(1)}}$ in (%
\ref{HMat}), computed numerically using a Floquet method, i.e. imposing
bosonic periodicity conditions $\psi (\theta +2\pi )=\psi (\theta )$. The
smallest critical value at $N\rightarrow \infty $ is in agreement with the
one reported in \cite{Moul}, whereas our values resulting from the smallest
zero for $N=5,7$ differ slightly from those reported in \cite{BijanMQR}. Our
values for $N\rightarrow \infty $ also agree precisely with those reported
in \cite{BenKal} when divided by a factor 2. We observe, that the rate of
convergence becomes very poor for larger values of the coupling constant and
even the lowest value requires fairly large values of $N$ to reach a good
precision.

A much better convergence can be obtained by taking the limit directly on
the level of the three-term recurrence relation. We easily see that for $%
N\rightarrow \infty $, $\zeta \rightarrow 0$\ with $g:=N\zeta <\infty $ and
the assumption that the coefficient functions remain finite, i.e. $%
\lim_{N\rightarrow \infty ,\zeta \rightarrow 0}\sigma _{n}=:\sigma _{n}^{M}$
and $\lim_{N\rightarrow \infty ,\zeta \rightarrow 0}c_{n}=:c_{n}^{M}$, the
relations (\ref{s1}), (\ref{c2}) and (\ref{c3}) reduce simply to%
\begin{eqnarray}
-g\sigma _{n-1}^{M}+4n^{2}\sigma _{n}^{M}+g\sigma _{n+1}^{M} &=&E\sigma
_{n}^{M},  \label{ei1} \\
-gc_{n-1}^{M}+4n^{2}c_{n}^{M}+gc_{n+1}^{M} &=&Ec_{n}^{M},  \label{ei2} \\
-2gc_{0}^{M}+4c_{1}^{M}+g\sigma _{2}^{M} &=&Ec_{1}^{M}.  \label{ei3}
\end{eqnarray}%
Rather than viewing these equations as recurrence relations we may also
interpret them as two separate eigenvalue equations for the infinite
matrices $\Xi $ and $\Theta $ with entries%
\begin{eqnarray}
\Xi _{i,j} &=&4i^{2}\delta _{i,i}+g\delta _{i,i+1}-g\delta _{i+1,i},\quad
\quad \qquad \qquad \quad \text{for }i,j\in \mathbb{N}, \\
\Theta _{i,j} &=&4i^{2}\delta _{i,i}+g\delta _{i,i+1}-g\delta
_{i+1,i}-g\delta _{i,2}\delta _{j,1},\quad \quad \text{for }i,j\in \mathbb{N}%
_{0},
\end{eqnarray}%
acting on the vectors $(\sigma _{1}^{M},\sigma _{2}^{M},\sigma
_{3}^{M},\ldots )$ and $(c_{0}^{M},c_{1}^{M},\sigma _{1}^{M},\ldots )$,
respectively. Truncating the matrices at a certain level, say $\ell $, we
compute the characteristic polynomials $\det (\Xi ^{\ell }-E\mathbb{I})$ and 
$\det (\Theta ^{\ell }-E\mathbb{I})$ together with their corresponding
discriminants $\Delta ^{\Xi }(g)$ and $\Delta ^{\Theta }(g)$ from (\ref%
{Sylvester}). The real zeros $g_{0}$ of these $\ell (\ell -1)$-order
polynomials define the exceptional points. We present our numerical results
in tables \ref{T2} and \ref{T3}.

\begin{table}[h]
\begin{center}
\begin{tabular}{c||c|c|c|c|c|c|c|}
$\ell $ & $g_{0}$ & $g_{0}$ & $g_{0}$ & $g_{0}$ & $g_{0}$ & $g_{0}$ & $g_{0}$
\\ \hline\hline
$2$ & $6.00000$ &  &  &  &  &  &  \\ \hline
$3$ & $6.97891$ &  &  &  &  &  &  \\ \hline
$4$ & $6.92848$ & $18.77091$ &  &  &  &  &  \\ \hline
$5$ & $6.92896$ & $24.29547$ &  &  &  &  &  \\ \hline
$6$ & $6.92895$ & $29.26843$ & $29.73862$ &  &  &  &  \\ \hline
$7$ & $6.92895$ & $30.10798$ & $34.30404$ &  &  &  &  \\ \hline
$8$ & $6.92895$ & $30.09660$ & $39.34849$ & $61.30789$ &  &  &  \\ \hline
$\vdots $ & $\vdots $ & $\vdots $ & $\vdots $ & $\vdots $ &  &  &  \\ \hline
$26$ & $6.928955$ & $30.09677$ & $69.59879$ & $125.4354$ & $130.5181$ & $%
197.6067$ & $251.2637$ \\ \hline
$27$ & $\mathbf{6.928955}$ & $\mathbf{30.09677}$ & $\mathbf{69.59879}$ & $%
\mathbf{125.4354}$ & $135.5878$ & $\mathbf{197.6067}$ & $261.6061$ \\ \hline
$26$ & $286.1126$ & $357.0076$ & $390.9532$ & $448.0887$ & $511.0770$ & $%
525.2021$ &  \\ \hline
$27$ & $\mathbf{286.1126}$ & $372.5999$ & $\mathbf{390.9532}$ & $468.8640$ & 
$512.1858$ & $551.0671$ &  \\ \hline
\end{tabular}%
\end{center}
\caption{Real zeros $g_{0}$ of the discriminant polynomials $\Delta ^{\Xi
}(g)$ converging to the exceptional points marked in bold.}
\label{T2}
\end{table}

\begin{table}[h]
\begin{center}
\begin{tabular}{c||c|c|c|c|c|c|c|}
$\ell $ & $g_{0}$ & $g_{0}$ & $g_{0}$ & $g_{0}$ & $g_{0}$ & $g_{0}$ & $g_{0}$
\\ \hline\hline
$2$ & $1.41421$ &  &  &  &  &  &  \\ \hline
$3$ & $1.46904$ &  &  &  &  &  &  \\ \hline
$4$ & $1.46877$ & $12.34951$ &  &  &  &  &  \\ \hline
$5$ & $1.46877$ & $17.88618$ &  &  &  &  &  \\ \hline
$6$ & $1.46877$ & $16.44658$ & $24.21371$ &  &  &  &  \\ \hline
$7$ & $1.46877$ & $16.47150$ & $29.27154$ &  &  &  &  \\ \hline
$8$ & $1.46877$ & $16.47116$ & $34.30396$ & $45.47616$ &  &  &  \\ \hline
$\vdots $ & $\vdots $ & $\vdots $ & $\vdots $ & $\vdots $ &  &  &  \\ \hline
$26$ & $1.46877$ & $16.47117$ & $47.80597$ & $95.47527$ & $125.4485$ & $%
159.4792$ & $239.8178$ \\ \hline
$27$ & $\mathbf{1.46877}$ & $\mathbf{16.47117}$ & $\mathbf{47.80597}$ & $%
\mathbf{95.47527}$ & $130.5181$ & $\mathbf{159.4792}$ & $\mathbf{239.8178}$
\\ \hline
$26$ & $240.9227$ & $336.4911$ & $341.4216$ & $427.3330$ & $449.3487$ & $%
498.9970$ &  \\ \hline
$27$ & $251.2637$ & $\mathbf{336.4911}$ & $357.0076$ & $448.0887$ & $%
449.5057 $ & $525.2659$ &  \\ \hline
\end{tabular}%
\end{center}
\caption{Real zeros $g_{0}$ of the discriminant polynomials $\Delta ^{\Theta
}(g)$ converging to the exceptional points marked in bold.}
\label{T3}
\end{table}

As a simple criterion we take the stabilization to a fixed value at some
given precision for two consecutive truncation levels. Taking for instance
the truncation levels to be $26$ and $27$ we have identified 14 exceptional
points marked in bold in our tables. All known values obtained agree with
those reported in the literature. Already at this level we find several new
values and it would be easy to extend these computation to higher truncation
levels being simply limited by computer power. Evidently solving the
characteristic polynomials for the energies and eigenvectors will provide a
good solution for the complete eigenvalue problem of $\mathcal{H}_{\text{Mat}%
}^{\Pi ^{(1)}}$, i.e. the eigenvectors and eigenfunctions, which we will
however not present here.

In a loose sense the discrete eigenvalue problem considered here is somewhat
similar in spirit to Znojil's \cite{Znojildis0,Znojildis1,Znojildis2}
discretized versions for non-Hermitian systems, although the discrete nature
of the problem results here from the coefficients in the expansion of the
eigenfunction rather than from the space on which they are defined. In a
slightly different approach the first levels of discretized version of the
complex Mathieu equation for some even solutions were also recently reported
in \cite{Ziener}.

\subsection{Factorization beyond the quantization level}

In our Ansatz we have truncated the Fourier series at a specific value,
which led to the quantization condition (\ref{EQ}). In fact this is not
necessary and instead we may also consistently consider the infinite series $%
\psi _{N\rightarrow \infty }^{s}(\theta )$ and $\psi _{N\rightarrow \infty
}^{c}(\theta )$ with $\zeta $ finite, i.e. here this is not to be understood
as a double scaling limit. Imposing then the quantization condition by hand
at some level all higher terms in the series above will automatically
vanish. This effect results from the factorization property of the
polynomials beyond that level in a similar fashion well known for the
Bender-Dunne polynomials \cite{Bender:1995rh}. We find here%
\begin{equation}
\sigma _{M+n}=\sigma _{M+1}Q_{n-1},\qquad \text{and\qquad }%
c_{L+m}=c_{L}Q_{m}\qquad \text{for }n\in \mathbb{N},m\in \mathbb{N}_{0},
\label{fact}
\end{equation}%
where we denoted $M=(N-1)/2$ and $L=(N+1)/2$. We will now provide
expressions for the polynomials $Q_{n}$. The factorization is easily
understood by noting first that $\gamma _{M}=0$, which simply has the effect
to reset the recurrence relations to the beginning, albeit with different
initial conditions. By inspection we see that the recurrence relations (\ref%
{canform}) from the level $M$ onwards simply reduces to $s_{M+1}=s_{M}$, $%
s_{M+2}=(1+\gamma _{M+1})s_{M}$, $s_{M+3}=(1+\gamma _{M+1}+\gamma
_{M+2})s_{M}$, $s_{M+4}=(1+\gamma _{M+1}+\gamma _{M+2}+\gamma _{M+3}+\gamma
_{M+1}\gamma _{M+3})s_{M}$, etc. Thus apart from the overall factor of $s_{M}
$, we simply recover the original solutions for $s_{n}$ with all functions $%
\gamma _{i}$ replaced by $\gamma _{i+M}$. Therefore we obtain the solutions%
\begin{equation}
s_{M+n}=s_{n}(\gamma _{i}\rightarrow \gamma _{i+M})s_{M}\qquad \text{for }%
n\in \mathbb{N}_{0}.
\end{equation}%
Using the first relation in (\ref{EQ}) this becomes 
\begin{equation}
\sigma _{M+n}=\sigma _{M+1}s_{n-1}(\gamma _{i}\rightarrow \gamma
_{i+M})\prod\limits_{k=M+1}^{M+n-1}\alpha _{k}=\sigma _{M+1}\hat{\sigma}_{n},
\end{equation}%
from which we may read of the polynomials $\hat{\sigma}_{n}$. For instance,
we have 
\begin{eqnarray}
\hat{\sigma}_{1} &=&1, \\
\hat{\sigma}_{2} &=&\alpha _{M+1}, \\
\hat{\sigma}_{3} &=&\alpha _{M+1}\alpha _{M+2}+\beta _{M+2}, \\
\hat{\sigma}_{4} &=&\alpha _{M+1}\alpha _{M+2}\alpha _{M+3}+\beta
_{M+2}\alpha _{M+3}+\beta _{M+3}\alpha _{M+1}.
\end{eqnarray}%
Likewise noting that $\lambda _{L}=0$, the recurrence relations (\ref{recq})
from that level onwards become $q_{L+1}=q_{L}$, $q_{L+2}=(1+\lambda
_{L+1})q_{L}$, etc. and therefore 
\begin{equation}
q_{L+n}=q_{n}(\lambda _{i}\rightarrow \lambda _{i+L})q_{L}\qquad \text{for }%
n\in \mathbb{N}_{0}.
\end{equation}%
The first relation in (\ref{cl}) then yields 
\begin{equation}
c_{L+n}=c_{L}q_{n}(\lambda _{i}\rightarrow \lambda
_{i+L})\prod\limits_{k=L}^{L+n-1}\alpha _{k}=c_{L}\hat{c}_{n},
\end{equation}%
from which we may identify the polynomials $\hat{c}_{n}$. We obtain for
example $\hat{c}_{0}=1$, $\hat{c}_{1}=\alpha _{L}$, $\hat{c}_{2}=\alpha
_{L}\alpha _{L+1}+\beta _{L+1}$, etc. We observe that $\hat{c}_{i}=\hat{%
\sigma}_{i+1}=Q_{i}$ for $i=0,1,2,\ldots $, thus establishing (\ref{fact}).

\subsection{Weakly orthogonal polynomials}

As seen in the previous subsection our univariate polynomials $\sigma _{n}(E)
$ and $c_{n}(E)$ enjoy similar properties as the Bender-Dunne polynomials
and are therefore expected to be also weakly orthogonal polynomials in $E$
in the sense defined in \cite{Finkel,Krajewska}. We briefly recall the
formalism and details of how this is achieved. As pointed out in \cite%
{Finkel}, according to Favard's theorem \cite{Favard} on\ three-term
recurrence relations of the general form%
\begin{equation}
\Phi _{n+1}=\left( E-a_{n}\right) \Phi _{n}-b_{n}\Phi _{n-1},  \label{recgen}
\end{equation}%
with $b_{n}=0$ for $n\leq 0$ and $b_{K}=0$, there exists always a linear
functional $\mathcal{L}$ acting on arbitrary polynomials $p$ as%
\begin{equation}
\mathcal{L}(p)=\int\nolimits_{-\infty }^{\infty }p(E)\omega (E)dE,  \label{L}
\end{equation}%
such that the polynomials $\Phi _{n}(E)$ are orthogonal%
\begin{equation}
\mathcal{L}(\Phi _{n}\Phi _{m})=\mathcal{L}(E\Phi _{n}\Phi
_{m-1})=N_{n}\delta _{nm}.  \label{ortho}
\end{equation}%
Thus the constants $N_{n}$ constitute the squared norms of $\Phi _{n}$. The
second equation in (\ref{ortho}) is a simple consequence of the first and (%
\ref{recgen}). The functional is unique \cite{Finkel} with the specific
initial conditions $\mathcal{L}(\Phi _{0}=1)=1$. The measure $\omega (E)$ is
given by the equations 
\begin{equation}
\omega (E)=\sum\limits_{k=1}^{K}\omega _{k}\delta (E-E_{k}),
\end{equation}%
where the energies $E_{k}$ are the $K$ roots of the polynomial $\Phi _{K}$
and the $K$ constants $\omega _{k}$ are determined by the $K$ equations%
\begin{equation}
\sum\limits_{k=1}^{K}\omega _{k}\Phi _{n}(E_{k})=\delta _{n0}\text{,}\qquad 
\text{for }n\in \mathbb{N}_{0}.  \label{om}
\end{equation}%
The squared norms $N_{n}$ may also be computed in an alternative way.
Multiplying (\ref{recgen}) by $\Phi _{n-1}$ and subsequently acting on it
with $\mathcal{L}$, assuming that it exists and satisfies (\ref{ortho}), we
obtain the simple homogeneous two-term relation%
\begin{equation}
N_{n}=b_{n}N_{n-1},\qquad \text{for }n\in \mathbb{N}.
\end{equation}%
Using the initial condition $N_{0}=1$ this is easily solved to%
\begin{equation}
N_{n}=\left\{ 
\begin{array}{ll}
\prod\limits_{k=1}^{n}b_{k},\qquad  & \text{for }1\leq n<K, \\ 
0, & \text{for }n\geq K.%
\end{array}%
\right.   \label{gamma}
\end{equation}%
Thus as already pointed out in \cite{Bender:1995rh} we may compute the
squared norms without the knowledge of the measure.

A further quantity which is easily computed with the knowledge of the
explicit form of $\mathcal{L}$ are the moment functionals as pointed out in 
\cite{Favard,Finkel}. They are defined as%
\begin{equation}
\mu _{n}:=\mathcal{L}(E^{n})=\sum\limits_{k=1}^{K}\int\nolimits_{-\infty
}^{\infty }E^{n}\omega _{k}\delta (E-E_{k})dE=\sum\limits_{k=1}^{K}\omega
_{k}E_{k}^{n},  \label{mu}
\end{equation}%
Once again also these quantities can be obtained alternatively from the
original polynomials without the knowledge of the $\omega _{k}$. Defining
the coefficients $\nu _{k}^{(n)}$ by the relation $\Phi
_{n}(E)=E^{n}-\sum\nolimits_{k=0}^{n-1}\nu _{k}^{(n)}E^{k}$ and acting on
this equation with $\mathcal{L}$ we obtain%
\begin{equation}
\mu _{n}=\sum\limits_{k=0}^{n-1}\nu _{k}^{(n)}\mu _{k}.  \label{mu2}
\end{equation}%
Evidently this allows for a recursive construction of all constants $\mu
_{n} $ when all polynomials $\Phi _{k}(E)$ with $k\leq n$ are known.

As a consistency check on the wavefunctions one may also compute%
\begin{equation}
\mathcal{L}\left[ \psi (\theta ,E)\right] =\sum\limits_{k=1}^{K}\int%
\nolimits_{-\infty }^{\infty }\psi (\theta ,E)\omega _{k}\delta
(E-E_{k})dE=\sum\limits_{k=1}^{K}\omega _{k}\psi (\theta ,E_{k}).
\end{equation}%
As a consequence of (\ref{om}) we expect for our wave functions (\ref{psis})
and (\ref{psic}) the simple expressions%
\begin{equation}
\mathcal{L}\left[ \psi _{N}^{s}(\theta ,E)\right] =-\sin (2\theta )e^{\frac{i%
}{2}\zeta \cos (2\theta )},\quad \text{and\quad }\mathcal{L}\left[ \psi
_{N}^{c}(\theta ,E)\right] =e^{\frac{i}{2}\zeta \cos (2\theta )}.  \label{xc}
\end{equation}

Let us now apply the above further to our system. For this purpose we
extract from the $\sigma _{n}(E)$ the denominator by introducing the new
polynomials $P_{n}(E)$ via the relation $\sigma _{n}=P_{n-1}\zeta
^{1-n}\prod\nolimits_{k=1}^{n-1}(N+1+2k)^{-1}$. The three-term recurrence
equation (\ref{s1}) then translates into a relation for the polynomials $%
P_{n}$%
\begin{equation}
P_{n+1}=\left[ E-\zeta ^{2}-4(n+1)^{2}\right] P_{n}+\zeta ^{2}\left[
N^{2}-(1+2n)^{2}\right] P_{n-1},\quad n\in \mathbb{N}_{0}.  \label{P}
\end{equation}%
Comparing with the generic form (\ref{recgen}) we identify $\Phi _{n}=P_{n}$
and the constants $a_{n}=$ $\zeta ^{2}+4(n+1)^{2}$, $b_{n}=\zeta ^{2}\left[
(1+2n)^{2}-N^{2}\right] $. Since $b_{M}=0$ our critical cut off level is $K=M
$. Therefore the squared norms result from (\ref{gamma}) to%
\begin{equation}
N_{n}=\left\{ 
\begin{array}{ll}
4^{n}\zeta ^{2n}\left( \frac{3}{2}-\frac{N}{2}\right) _{n}\left( \frac{3}{2}+%
\frac{N}{2}\right) _{n},\qquad  & \text{for }1\leq n\leq (N-3)/2, \\ 
0, & \text{for }n>(N-3)/2,%
\end{array}%
\right.   \label{gn}
\end{equation}%
where $(a)_{n}:=\Gamma \left( a+n\right) /\Gamma \left( a\right) $ denotes
the Pochhammer symbol.

For concrete values of $N$ these expressions are easily computed. Taking for
instance $N=7$ our general expressions for $\sigma _{n}$ yield 
\begin{eqnarray}
P_{0}(E) &=&1,  \label{PP} \\
P_{1}(E) &=&E-\zeta ^{2}-4,  \notag \\
P_{2}(E) &=&E^{2}-2E\left( \zeta ^{2}+10\right) +\zeta ^{4}+60\zeta ^{2}+64,
\notag \\
P_{3}(E) &=&E^{3}-E^{2}\left( 3\zeta ^{2}+56\right) +E\left( 3\zeta
^{4}+176\zeta ^{2}+784\right) -\zeta ^{6}-120\zeta ^{4}-2320\zeta ^{2}-2304,
\notag \\
P_{4}(E) &=&\left( E-\zeta ^{2}-64\right) P_{3}(E),  \notag \\
P_{5}(E) &=&\left[ E^{2}-2E\left( \zeta ^{2}+82\right) +\zeta ^{4}+132\zeta
^{2}+6400\right] P_{3}(E).  \notag
\end{eqnarray}%
We observe the typical factorization above the quantization level, that is $%
M=3$ in this case. The roots of the polynomial $P_{3}(E)$ were computed
before in (\ref{E7}). Assigning here $E_{1}:=E_{7}^{s,-2}$, $%
E_{2}:=E_{7}^{s,2}$ and $E_{3}:=E_{7}^{s,0}$ we solve the three equations (%
\ref{om}) for the constants $\omega _{k}$. We find%
\begin{equation}
\omega _{1}=\varkappa _{2},\qquad \omega _{2}=\varkappa _{-2},\qquad \omega
_{3}=\varkappa _{0}.  \label{ome}
\end{equation}%
where we abbreviated%
\begin{equation*}
\varkappa _{\ell }:=\frac{2\left( 49-12\zeta ^{2}\right) ^{2}+3\left(
48-7\zeta ^{2}\right) e^{-\frac{i\pi \ell }{3}}\Omega ^{2}+22\left( 12\zeta
^{2}-49\right) e^{\frac{i\pi \ell }{3}}\Omega +2e^{\frac{i\pi \ell }{3}%
}\Omega ^{4}-22\Omega ^{3}}{6\left[ \left( 49-12\zeta ^{2}\right)
^{2}+\left( 49-12\zeta ^{2}\right) e^{-\frac{i\pi \ell }{3}}\Omega ^{2}+e^{%
\frac{i\pi \ell }{3}}\Omega ^{4}\right] },
\end{equation*}%
with $\Omega $ as defined after (\ref{E7}). This gives a complicated
measure, but when computing the integrals the result simplifies to%
\begin{eqnarray}
\mathcal{L}(P_{0}^{2}) &=&\omega _{1}+\omega _{2}+\omega _{3}=1=N_{0},
\label{g1} \\
\mathcal{L}(P_{1}^{2}) &=&\omega _{1}P_{1}^{2}(E_{1})+\omega
_{2}P_{1}^{2}(E_{2})+\omega _{3}P_{1}^{2}(E_{3})=-40\zeta ^{2}=N_{1}, \\
\mathcal{L}(P_{2}^{2}) &=&\omega _{1}P_{2}^{2}(E_{1})+\omega
_{2}P_{2}^{2}(E_{2})+\omega _{3}P_{2}^{2}(E_{3})=960\zeta ^{4}=N_{2},
\label{g2} \\
\mathcal{L}(P_{m}^{2}) &=&0,\qquad \text{for }m\geq 3, \\
\mathcal{L}(P_{n}P_{m}) &=&0,\qquad \text{for }m\neq n\text{; }n,m\in 
\mathbb{N}_{0}.  \label{g5}
\end{eqnarray}%
where (\ref{g1}) - (\ref{g2}) agree exactly with the expressions computed
from (\ref{gn}). In (\ref{g5}) especially $\mathcal{L}(P_{1}P_{2})$ is a
non-trivial check. We also recover the first expression in (\ref{xc}) when
using (\ref{ome}).

Employing next the formula (\ref{mu}) together with (\ref{ome}) we compute
the first momentum functionals to 
\begin{eqnarray}
\mu _{0} &=&1, \\
\mu _{1} &=&\zeta ^{2}+4, \\
\mu _{2} &=&\zeta ^{4}-32\zeta ^{2}+16, \\
\mu _{3} &=&\zeta ^{6}-108\zeta ^{4}-912\zeta ^{2}+64, \\
\mu _{4} &=&\zeta ^{8}-224\zeta ^{6}-1184\zeta ^{4}-17024\zeta ^{2}+256.
\end{eqnarray}%
We easily verify that these expressions also satisfy the relation (\ref{mu2}%
) together with (\ref{PP}).

Turning now to our second set of recurrence relations (\ref{c2}), (\ref{c3})
we proceed in a similar fashion and extract the denominator $c_{n}(E)$ by
introducing the new polynomials $\hat{P}_{n}(E)$ via $c_{n}=\hat{P}_{n}\zeta
^{-n}\prod\nolimits_{k=1}^{n}(N+2k-1)^{-1}$. The three-term recurrence
relations (\ref{c2}), (\ref{c3}) then becomes%
\begin{eqnarray}
\hat{P}_{n+1} &=&\left( E-\zeta ^{2}-4n^{2}\right) \hat{P}_{n}+\zeta ^{2}%
\left[ N^{2}-(2n-1)^{2}\right] \hat{P}_{n-1},\quad n\in \mathbb{N}%
_{0}\backslash \{1\}, \\
\hat{P}_{2} &=&\left( E-\zeta ^{2}-4\right) \hat{P}_{1}+2\zeta ^{2}\left(
N^{2}-1\right) \hat{P}_{0}.
\end{eqnarray}%
Now we identify $\Phi _{n}=\hat{P}_{n}$ and the constants $\hat{a}_{n}=$ $%
\zeta ^{2}+4n^{2}$ for $n\in \mathbb{N}_{0}$, $\hat{b}_{1}=$ $2\zeta
^{2}\left( N^{2}-1\right) ,\hat{b}_{n}=\zeta ^{2}\left[ (2n-1)^{2}-N^{2}%
\right] $ for $n\in \mathbb{N}_{0}\backslash \{1\}$. Since $b_{L}=0$ our
critical cut off level results to $K=L$. The squared norms computed from (%
\ref{gamma}) are%
\begin{equation}
\hat{N}_{n}=\left\{ 
\begin{array}{ll}
2^{2n+1}\zeta ^{2n}\left( \frac{1}{2}-\frac{N}{2}\right) _{n}\left( \frac{1}{%
2}+\frac{N}{2}\right) _{n},\qquad  & \text{for }1\leq n\leq (N-1)/2, \\ 
0, & \text{for }n>(N-1)/2.%
\end{array}%
\right.   \label{Nhat}
\end{equation}%
As an example we take now $N=5$ in our general expressions for $c_{n}$
obtaining 
\begin{eqnarray}
\hat{P}_{0}(E) &=&1, \\
\hat{P}_{1}(E) &=&E-\zeta ^{2},  \notag \\
\hat{P}_{2}(E) &=&E^{2}-2E\left( \zeta ^{2}+2\right) +\zeta ^{2}\left( \zeta
^{2}+52\right) ,  \notag \\
\hat{P}_{3}(E) &=&E^{3}+E^{2}\left( -3\zeta ^{2}-20\right) +E\left( 3\zeta
^{4}+104\zeta ^{2}+64\right) -\zeta ^{2}\left( \zeta ^{4}+84\zeta
^{2}+832\right)   \notag \\
\hat{P}_{4}(E) &=&\left( E-\zeta ^{2}-36\right) \hat{P}_{3}(E),  \notag \\
\hat{P}_{5}(E) &=&\left[ E^{2}-2E\left( \zeta ^{2}+50\right) +\zeta
^{4}+76\zeta ^{2}+2304\right] \hat{P}_{3}(E).  \notag
\end{eqnarray}%
As expected the polynomials factorize again after the quantization level.
Assigning now the previously computed roots of $\hat{P}_{3}(E)$, see (\ref%
{E5}), as $E_{1}=E_{5}^{c,-2}$, $E_{2}=E_{5}^{c,2}$ and $E_{3}=E_{5}^{c,0}$
we solve the three equations (\ref{om}) for the constants $\omega _{k}$. We
compute%
\begin{equation}
\hat{\omega}_{1}=\hat{\varkappa}_{2},\qquad \hat{\omega}_{2}=\hat{\varkappa}%
_{-2},\qquad \hat{\omega}_{3}=\hat{\varkappa}_{0}.  \label{omehat}
\end{equation}%
where we defined the function%
\begin{equation*}
\hat{\varkappa}_{\ell }:=\frac{\left( 13-12\zeta ^{2}\right) ^{2}+\left(
12-15\zeta ^{2}\right) e^{-\frac{i\pi k}{3}}\hat{\Omega}^{2}+\left( 60\zeta
^{2}-65\right) e^{\frac{i\pi k}{3}}\hat{\Omega}+e^{\frac{i\pi k}{3}}\hat{%
\Omega}^{4}-5\hat{\Omega}^{3}}{3\left( 13-12\zeta ^{2}\right) ^{2}+\left(
39-36\zeta ^{2}\right) e^{-\frac{i\pi k}{3}}\hat{\Omega}^{2}+3e^{\frac{i\pi k%
}{3}}\hat{\Omega}^{4}},
\end{equation*}%
with $\hat{\Omega}$ as introduced after (\ref{E5}). Next we use this measure
to evaluate%
\begin{eqnarray}
\mathcal{L}(\hat{P}_{0}^{2}) &=&\hat{\omega}_{1}+\hat{\omega}_{2}+\hat{\omega%
}_{3}=1=\hat{N}_{0}, \\
\mathcal{L}(\hat{P}_{1}^{2}) &=&\hat{\omega}_{1}P_{1}^{2}(E_{1})+\hat{\omega}%
_{2}P_{1}^{2}(E_{2})+\hat{\omega}_{3}P_{1}^{2}(E_{3})=-48\zeta ^{2}=\hat{N}%
_{1}, \\
\mathcal{L}(\hat{P}_{2}^{2}) &=&\hat{\omega}_{1}P_{2}^{2}(E_{1})+\hat{\omega}%
_{2}P_{2}^{2}(E_{2})+\hat{\omega}_{3}P_{2}^{2}(E_{3})=768\zeta ^{4}=\hat{N}%
_{2}, \\
\mathcal{L}(\hat{P}_{m}^{2}) &=&0,\qquad \text{for }m\geq 3, \\
\mathcal{L}(\hat{P}_{n}\hat{P}_{m}) &=&0,\qquad \text{for }m\neq n\text{; }%
n,m\in \mathbb{N}_{0}.
\end{eqnarray}%
We verify that the results for $\hat{N}_{1}$ and $\hat{N}_{2}$ obtained by
direct integration agree precisely with those computed from the expression
in (\ref{Nhat}). In addition, we recover the second identity in (\ref{xc})
from the integration with (\ref{omehat}).

Using formula (\ref{mu}) together with (\ref{omehat}) we compute the first
momentum functionals to 
\begin{eqnarray}
\hat{\mu}_{0} &=&1, \\
\hat{\mu}_{1} &=&\zeta ^{2}, \\
\hat{\mu}_{2} &=&\zeta ^{4}-48\zeta ^{2}, \\
\hat{\mu}_{3} &=&\zeta ^{6}-144\zeta ^{4}-192\zeta ^{2}, \\
\hat{\mu}_{4} &=&\zeta ^{8}-288\zeta ^{6}+2304\zeta ^{4}-768\zeta ^{2},
\end{eqnarray}%
which again agree with those evaluated from (\ref{mu2}).

\section{Conclusions and outlook}

We have demonstrated that $E_{2}$-quasi-exact solvability allows for a
systematic construction of solutions to the eigenvalue problem for models
that can be cast into that form. For the non-Hermitian models presented
here, $\mathcal{H}_{N}$, but especially $\mathcal{H}_{\text{Mat}}$, the
analysis seems to be more natural, easier and transparent than in a $sl_{2}(%
\mathbb{C})$-setting. A sharper comparison between the two schemes can of
course be developed. Obviously the above analysis is not restricted to
non-Hermitian systems, which were selected as they are of current interest,
especially in the context of optics \cite%
{Muss,MatMakris,Guo,OPMidya,MatHugh,MatHughEva,MatLongo}. There is large
scope to search for other models that are $E_{2}$-quasi-exactly solvable by
modifying the vector spaces (\ref{V1}) and (\ref{V2}) by adjusting them for
Hermitian models or systems possessing different types of $\mathcal{PT}$%
-symmetry as identified in \cite{DFM}.

In the analysis of the complex Mathieu equation we have mainly focused here
on the identification of the exceptional points. In that regard we
demonstrated that the double scaling limit for $\mathcal{H}_{N}$ level by
level provides a very good qualitative picture for the Mathieu system, but
to achieve quantitative precision requires to compute very high levels.
Alternatively it is much easier and efficient to carry out the limit
directly for the three-term recurrence relation and view that equation as an
eigenvalue equation. One is still left with an entirely algebraic setting
and convergence is achieved very fast. We have reported several hitherto
unknown exceptional points. The latter approach is also far more efficient
than the Floquet method, which for the problem discussed becomes very
unstable for large values of $\zeta $ .

\newif\ifabfull\abfulltrue

\end{document}